\documentclass[aps,prb,twocolumn,groupedaddress,
showpacs,amsmath,amssymb]{revtex4}

\usepackage{graphicx}
\usepackage{dcolumn}
\usepackage{bm}

\begin{document}

\title{Normal-state conductivity in underdoped La$_{2-x}$Sr$_x$CuO$_4$
thin films: Search for nonlinear effects related to collective stripe
motion}

\author{A. N. Lavrov}
\author{I. Tsukada}
\author{Yoichi Ando}

\affiliation{Central Research Institute of Electric Power Industry,
Komae, Tokyo 201-8511, Japan}

\date{\today}

\begin{abstract}

We report a detailed study of the electric-field dependence of the
normal-state conductivity in La$_{2-x}$Sr$_x$CuO$_4$ thin films for two
concentrations of doped holes, $x=0.01$ and 0.06, where formation of
diagonal and vertical charged stripes was recently suggested. In order to
elucidate whether high electric fields are capable of depinning the
charged stripes and inducing their collective motion, we have measured
current-voltage characteristics for various orientations of the electric
field with respect to the crystallographic axes. However, even for the
highest possible fields ($\sim 1000$ V/cm for $x=0.01$ and $\sim 300$ V/cm
for $x=0.06$) we observed no non-linear-conductivity features except for
those related to the conventional Joule heating of the films. Our analysis
indicates that Joule heating, rather than collective electron motion, may
also be responsible for the non-linear conductivity observed in some other
2D transition-metal oxides as well. We discuss that a possible reason why
moderate electric fields fail to induce a collective stripe motion in
layered oxides is that fairly flexible and compressible charged stripes
can adjust themselves to the crystal lattice and individual impurities,
which makes their pinning much stronger than in the case of conventional
rigid charge-density waves.

\end{abstract}

\pacs{74.25.Fy, 74.72.Dn, 71.45.Lr, 72.20.Ht}

\maketitle

\section{Introduction}

The parent compounds of high-$T_c$ cuprates are known to be correlated
Mott insulators that become metallic and superconducting (SC) upon doping
with charge carriers; the mechanism of this evolution, however, still
remains a mystery. One of the possible pictures is that the doped holes
segregate, instead of being homogeneously distributed, and establish an
array of microscopic conducting channels (charged stripes) embedded in the
insulating matrix.\cite{stripe_rev,NdSr,diag_ver,mobility,anis} In fact,
these conducting channels reduce the penalty for disrupting the correlated
insulating state, and allow even a few holes to move through a Mott
insulator. Owing to the long-range Coulomb interaction, the hole-rich
channels tend to order into a fairly periodic pattern, reminiscent of the
charge-density wave (CDW) in quasi-1D conductors.\cite{CDW_rev} Periodic
charge-density modulations have indeed been found in some of the cuprate
compounds, \cite{stripe_rev,NdSr,1eighth} giving support to the stripe
picture.

Despite the resemblance, the charged stripes differ from conventional CDW
both in properties and in the mechanism driving their formation: A
conventional CDW is governed by the Fermi-surface instability of a metal
and results in opening of a gap exactly at the Fermi level,\cite{CDW_rev}
while charged stripes stem from the tendency of doped holes to avoid
localization, and do not require such gap formation.\cite{stripe_rev} The
absence of a gap at the Fermi level allows the stripes to be conducting,
and also makes them compressible -- the hole filling of stripes as well as
the distance between them should be readily variable. Consequently, the
conducting stripes may well be flexible and fluctuating in contrast to
rather rigid CDW.

Apparently, those strong fluctuations make the stripes in cuprates quite
elusive, causing many experiments aimed at observing the stripes to fail.
Although one can easily find evidence for microscopic charge
inhomogeneity, no static stripe ordering is observed unless a strong
collective pinning (commensurability effects at $1/8$ filling, structural
distortions in La$_{2-x-y}$Nd$_y$Sr$_x$CuO$_4$, {\it etc.}) fixes the
position and orientation of the stripes in CuO$_2$ planes, making them
visible for diffraction techniques.\cite{stripe_rev,NdSr,1eighth} This
naturally casts doubts on whether the stripes are inherent in cuprates and
ultimately relevant for high-$T_c$ superconductivity, or the observed
stripy superstructures are just a side effect caused by lattice
instabilities. Even more challenging is to find out what are the new
qualitative features that the charged stripes are bringing about.

Upon selecting experiments to clarify the role of stripes, one may consult
how the existence of collective electron states has been substantiated in
other systems, particularly when diffraction methods were incapable of
giving a conclusive evidence. In the field of inorganic quasi-1D
compounds, the key experiments that have led the CDW picture to triumph,
and have ultimately convinced researchers that they are dealing with a
truly collective state, were (i) observation of a sharp threshold electric
field in conductivity, corresponding to the onset of coherent CDW sliding,
and (ii) observation of a ``narrow-band noise'' induced by the motion of a
washboard-like CDW over defects.\cite{CDW_thr} The transport measurements
were certainly indispensable for 2D electron systems (2DES) in
heterostructures, where a conducting layer is buried deep in the crystal
and diffraction methods can hardly be used. A variety of collective
electron states including stripe, ``bubble'', and Wigner-crystal phases
expected \cite{2DEG_the} to be realized in 2DES were also documented by
observations of threshold conduction and narrow-band
noise.\cite{Wign,2DEG1} Another class of experiments is related to
qualitatively new features introduced by the collective state to the
single-particle transport. The most fascinating among those is the
observation of a large resistivity anisotropy which spontaneously develops
in seemingly isotropic 2DES at low temperatures, and whose orientation can
be switched by a magnetic field. \cite{2DEG2}

In cuprates, the qualitative evidence for conducting stripes, collected
thus far from transport measurements, is limited to a spontaneous (or
field-induced) in-plane resistivity anisotropy that develops at low
temperatures, \cite{anis} in striking resemblance to 2DES. A suppression
of the Hall resistivity in the static-stripe system
La$_{2-x-y}$Nd$_y$Sr$_x$CuO$_4$, initially considered as a clear evidence
for 1D transport, \cite{Hall} has been later understood as coming from a
tricky cancellation of the hole and electron terms, which may or may not
be related to the 1D hole motion; in fact, such Hall-resistivity
suppression is a rare exception among cuprates. \cite{mobility,Hall2} It
might sound surprising, but such key features as narrow-band noise or
threshold conductivity have never been seriously looked for in high-$T_c$
cuprates, though non-linear conduction has been observed in ladder
cuprates. \cite{ladder,ladder2} This is partly because of a common wisdom
which tells us that the CDW (or stripe) sliding is hardly possible in
2D/3D systems because of too strong pinning. However, this understanding
has been challenged recently by a number of papers reporting spectacular
non-linear conduction in layered nickel and manganese oxides,
\cite{Nature,Science,LaSrNiO,film,film_perp} which has been attributed to
the collective charge motion and the collapse of the charge-ordered state.
If this interpretation is actually correct, one may look for similar
stripe-sliding effects in cuprates \cite{cuprates} which, if found, would
finally clarify the electronic state underlying the high-$T_c$
superconductivity.

In this study, we search for non-linear conductivity features in the most
promising system La$_{2-x}$Sr$_x$CuO$_4$ (LSCO), where static and dynamic
stripes of different topology have been observed by neutron
scattering.\cite{stripe_rev,NdSr,diag_ver} The compositions $x=0.01$ and
$x=0.06$ are chosen as representing the ``diagonal'' and ``vertical''
stripe states\cite{diag_ver} (Fig. 1). In order to minimize the Joule
heating, we prepare LSCO thin films patterned into narrow bridges, which
allows us to perform current-voltage-characteristics measurements up to
electric fields of 100-1000 V/cm. The bridges are formed along several
crystallographic directions, making possible the field application along
or transverse to the expected stripe direction. However, up to the highest
electric fields we observe no non-linear conductivity features other than
those related to the Joule heating. This indicates that the electric-field
energy integrated over the correlated stripe volume (if the stripe picture
is actually appropriate for cuprates) is still too weak to overcome
pinning and to drive the stripe sliding. Consequently, the correlated
volume for the stripe ordering in cuprates appear to be much smaller and
the stripe pinning to be much stronger than in conventional CDW
systems,\cite{CDW_theo} and moderate electric fields may never be able to
induce the stripe sliding. Furthermore, a simple analysis of Joule heating
shows that electric fields of the order of 100-1000 V/cm represent a
characteristic threshold for many ``insulating'' transition-metal oxides.
For fields above this threshold, the samples {\it must} show non-linear
conductivity and switching phenomena which, however, are related neither
to collective charge motion, nor to other electronic peculiarities, but
are caused simply by overheating. This calls for more caution in
interpreting numerous observations of the non-linear conduction in
transition-metal oxides.

\section{Experimental details}

A difficult problem one inevitably faces upon measuring the
high-electric-field characteristics is the Joule overheating of samples.
Often used simultaneous current and voltage limitations
\cite{Nature,Science,LaSrNiO,film,film_perp} merely result in stabilizing
an inhomogeneous state, e.g., composed of conducting
filaments,\cite{Bonch,hot_spot} which hides the intrinsic behavior.
Another approach is to employ short-pulse technique; however, to reduce
overheating to a reasonable level, the pulses should be as short as 1
$\mu$s or less in most cases.\cite{mesa_heat,hot_spot} For CDW systems
possessing huge dielectric constants and strong frequency dependence of
the conductivity, \cite{CDW_rev} such short-pulse measurements would give
data that have nothing to do with the dc conduction. Apparently, the only
effective approach is to reduce the size of samples in order to decrease
the produced heat and to ease the heat removal. This miniaturization is
naturally limited by the characteristic correlation length of the ordered
state under investigation, when the surface pinning and size effects
become important.\cite{CDW_size} For high-electric-field measurements,
therefore, we chose thin-film samples, and employed a conventional dc
four-probe method.

\begin{figure}[!tb]
\leftskip18pt
\includegraphics*[width=16.5pc]{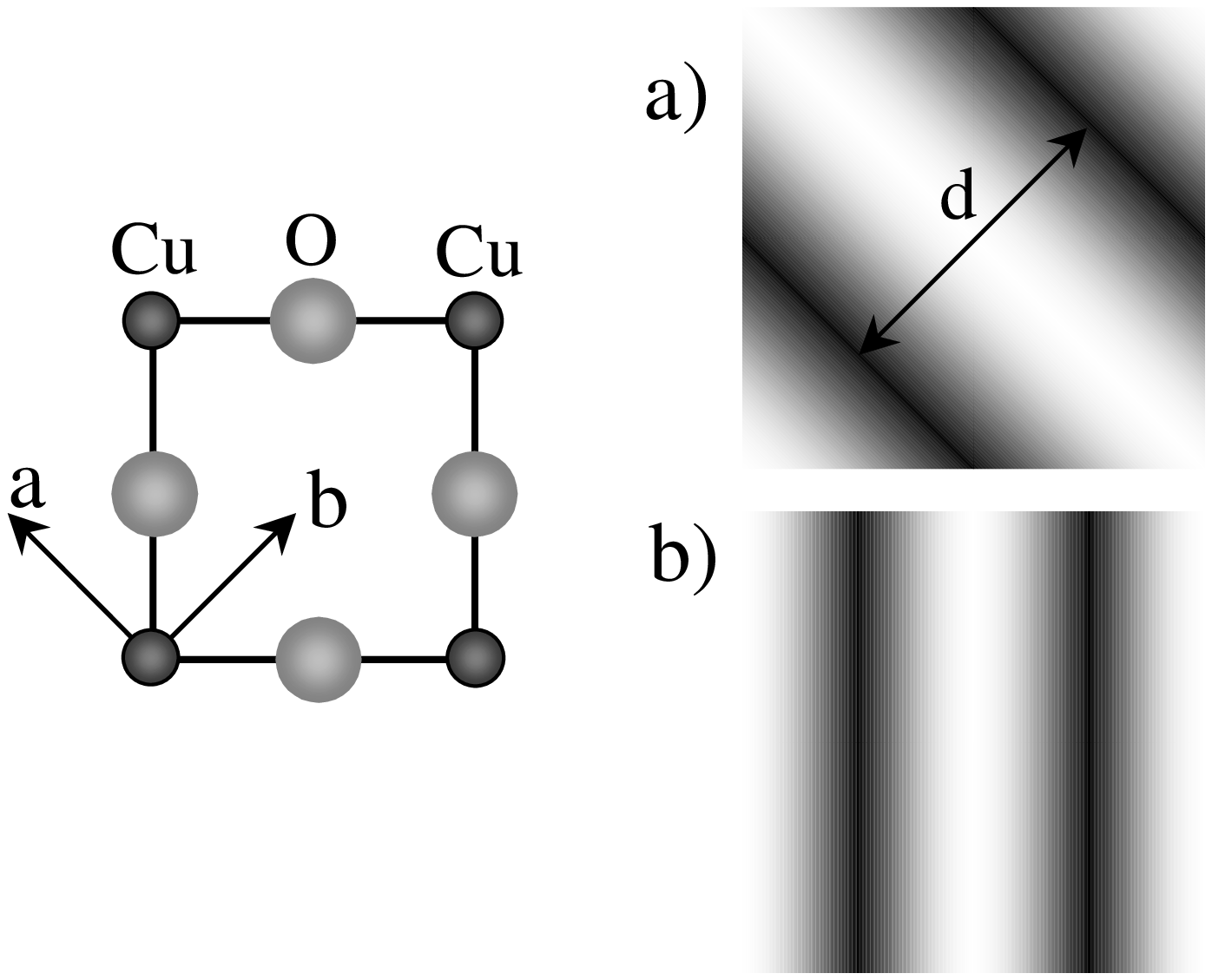}
\caption{Schematic picture of the CuO$_2$ plane, and an expected topology of
the charge modulation: a) diagonal stripes with the periodicity $d$
running along the orthorhombic $a$ axis (Cu-Cu direction); b) vertical
stripes running along the Cu-O-Cu directions.}
\label{sketch}
\end{figure}

Epitaxial La$_{2-x}$Sr$_x$CuO$_{4+\delta}$ films with $x=0.01$ and 0.06
were prepared by a conventional pulsed-laser deposition technique. During
the growth, the temperature of the substrate was set at 800 $\sim
830^{\circ}$C, and the oxygen pressure was kept around 4 Pa. An important
point was a proper choice of substrates, in order to minimize any unwanted
film distortion induced by the lattice mismatch. Since LSCO with the
$x=0.01$ composition was orthorhombic and was expected to possess
``diagonal'' stripes running along one of the orthorhombic Cu-Cu
directions (Fig. 1), we selected orthorhombic YAlO$_3$ (YAP) substrates
for growing $x=0.01$ LSCO films. In doing so, we intended to obtain films
with perfectly aligned crystallographic axes, and thus possessing a {\it
unidirectional} stripe structure. We indeed succeeded in growing untwinned
La$_{1.99}$Sr$_{0.01}$CuO$_4$ films on the (001) surface of YAP, where the
in-plane orientation LSCO [100] was parallel to YAP [100], according to
the x-ray diffraction. LSCO $x=0.06$ films, which were expected to have
``vertical'' stripes, were prepared on the (100) surface of SrTiO$_3$
(STO) and (001) surface of LaSrAlO$_4$ (LSAO) substrates. Both STO and
LSAO have a slight lattice mismatch with LSCO, yet this mismatch is of
different signs; thus, the LSCO films deposited on these substrates are
subject to an expansive and compressive in-plane strain,
respectively.\cite{strain} Since the epitaxial strain can easily affect
the stripe pinning, as it does with the superconducting transition
temperature, we use films both on STO and LSAO for a comparative study of
the current-voltage characteristics.

The thickness of prepared La$_{2-x}$Sr$_x$CuO$_{4+\delta}$ films was
determined to be $\approx 1200$ {\AA} and 2400 {\AA} for $x=0.01$ films,
and $\approx 1000$ {\AA} for $x= 0.06$ films (a piece of film was
dissolved in acid and the amount of material was measured by the
inductively-coupled plasma spectrometry). Each film was patterned into
narrow, $\sim20-50$ $\mu$m, bridges aligned along the Cu-Cu or Cu-O-Cu
directions, using photolithography. Electric contacts were made by gold
paint with subsequent annealing in pure helium (for $x=0.01$ films) and in
air (for $x= 0.06$ films), following the heat treatment procedure
developed for bulk crystals, \cite{anneal} which is required to establish
the oxygen stoichiometry $\delta = 0$.

The current-voltage characteristics were determined by applying a small
low-frequency ac modulation voltage to the sample while a dc bias voltage
was slowly swept, and measuring the differential conductance $dI/dV$. Upon
measurements, the substrate with sample was attached to a copper block,
whose temperature $T_{\text{base}}$ was stabilized with an accuracy better
than 0.01 K. The angular dependence of the magnetoresistance (MR) was
measured by rotating the sample at the fixed temperature and magnetic
field.

\section{Results and discussion}
\subsection{Resistivity and magnetoresistance}

\begin{figure}[!t]
\leftskip18pt
\includegraphics*[width=18pc]{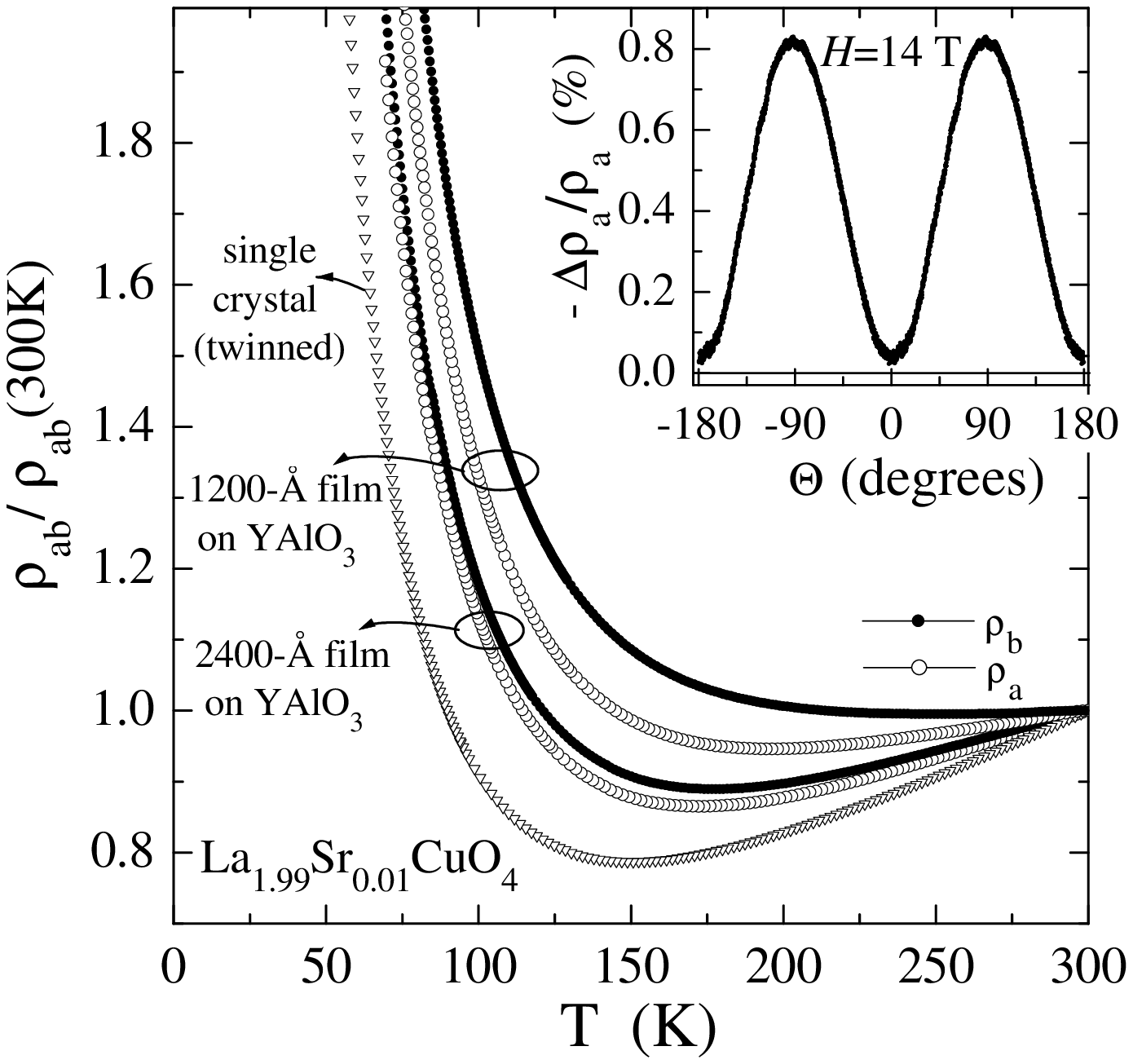}
\caption{Normalized resistivity of LSCO ($x=0.01$) films
deposited on YAP in comparison with single-crystal data from
Ref.[\onlinecite{mobility}]. 4-probe measurements are done on narrow
bridges formed along the $a$ or $b$ axis. Inset: angular dependence of the
magnetoresistance measured at 100 K upon rotating the 14-T magnetic field
within the $ab$ plane (parallel to the film).}
\label{R_YAP}
\end{figure}

It is well known that crystal defects and strains, including those induced
by a mismatch with the substrate, grain boundaries or surface effects, can
easily pin the CDW/stripe structure, preventing it from
sliding.\cite{CDW_rev,CDW_size} It is important, therefore, to obtain
thin-film samples with properties not much different from those of
high-quality single crystals. In the case of light doping ($x=0.01$), we
have succeeded in preparing LSCO films on YAlO$_3$ with the resistivity
behavior quite similar to that of single crystals,\cite{mobility} but the
film thickness had to be kept above $1000$ \AA~~(Fig. 2). A resistivity
upturn appears at somewhat higher temperatures in thinner films,
indicating easier localization of holes and larger disorder.

According to the neutron-scattering data, compositions with $x\leq 0.05$
possess unidirectional stripes running along the orthorhombic $a$ axis.
\cite{diag_ver} In order to compare $I-V$ characteristics along and
transverse to the stripes, one needs a single-crystalline film with
uniform orientation of the orthorhombic $a$ and $b$ axes. Previous studies
\cite{suscept,MR_LSCO} of detwinned LSCO $x=0.01$ single crystals have
revealed a strong in-plane anisotropy of the susceptibility and
magnetoresistance: When a magnetic field is applied along the $ab$ plane,
only the $b$ component of the field affects the spin and stripe structure
and causes magnetoresistance.\cite{MR_LSCO} Our LSCO films actually
demonstrate a clear $\sin^2\theta$ angular dependence of the MR (inset of
Fig. 2), indicating the crystallographic axes are perfectly aligned.

\begin{figure}[!t]
\leftskip25pt
\includegraphics*[width=16pc]{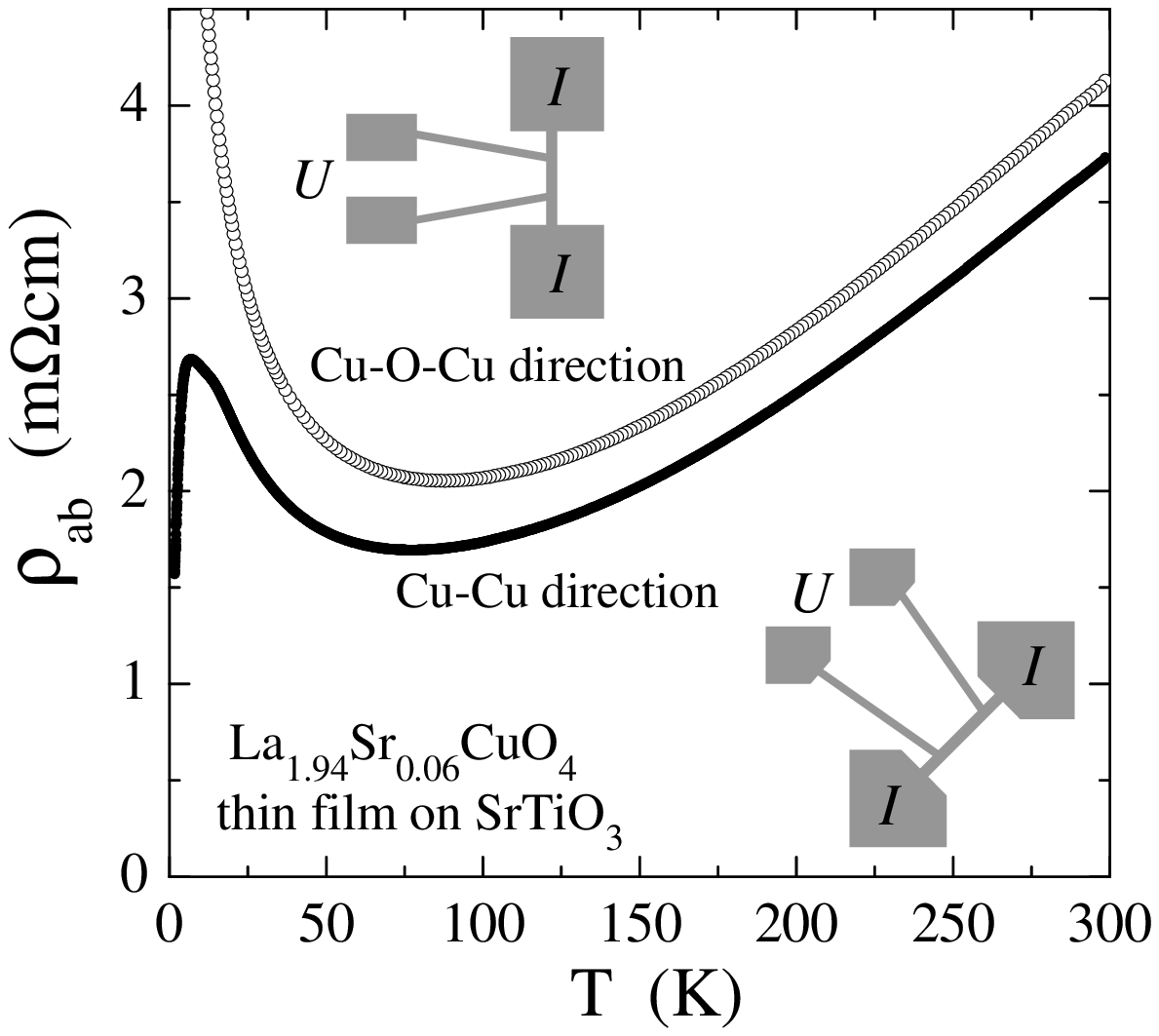}
\caption{Resistivity of 1000-\AA ~LSCO ($x=0.06$) films
deposited on STO. Insets illustrate the arrangement of narrow bridges
along the Cu-Cu or Cu-O-Cu directions.}
\label{R_STO}
\end{figure}

The composition $x=0.06$ is located just on the verge of the
superconductivity, where the stripes are also reported to change their
orientation from ``diagonal'', that is being parallel to the orthorhombic
axes, to the ``vertical'' one. Correspondingly, to check all possible
geometries, we prepared bridges directed along the ``diagonal'', Cu-Cu,
and ``vertical'', Cu-O-Cu, directions (insets of Fig. 3). The resistivity
behavior of LSCO $x=0.06$ films deposited on SrTiO$_3$ and LaSrAlO$_4$
substrates (Figs. 3 and 4) demonstrates that they are of high quality: The
resistivity values are close to those observed in the best single
crystals,\cite{mobility} $\rho_{ab}(300$K$)\approx 2$ m$\Omega$cm, and the
linear fitting of the high-temperature resistivity ($\rho(T)=\rho_0+AT$)
gives $\rho_0\approx 0$, indicating negligible impurity scattering. The
films however show some dispersion in properties. For example, in films on
STO (Fig. 3), both resistivity and $T_c$ vary, indicating slightly
different doping levels. In films on LSAO (Fig. 4), the doping seems to be
the same for all films, as follows from the position of the SC transition
(inset of Fig. 4), yet the resistivity does vary. Nevertheless, for this
particular composition on the border of the superconductivity region
($x=0.06$), one can hardly achieve better homogeneity: SC transitions
depicted in the inset of Fig. 4 are already among the narrowest ever
reported for single crystals or thin film.\cite{mobility}

\begin{figure}[!t]
\leftskip18pt
\includegraphics*[width=17pc]{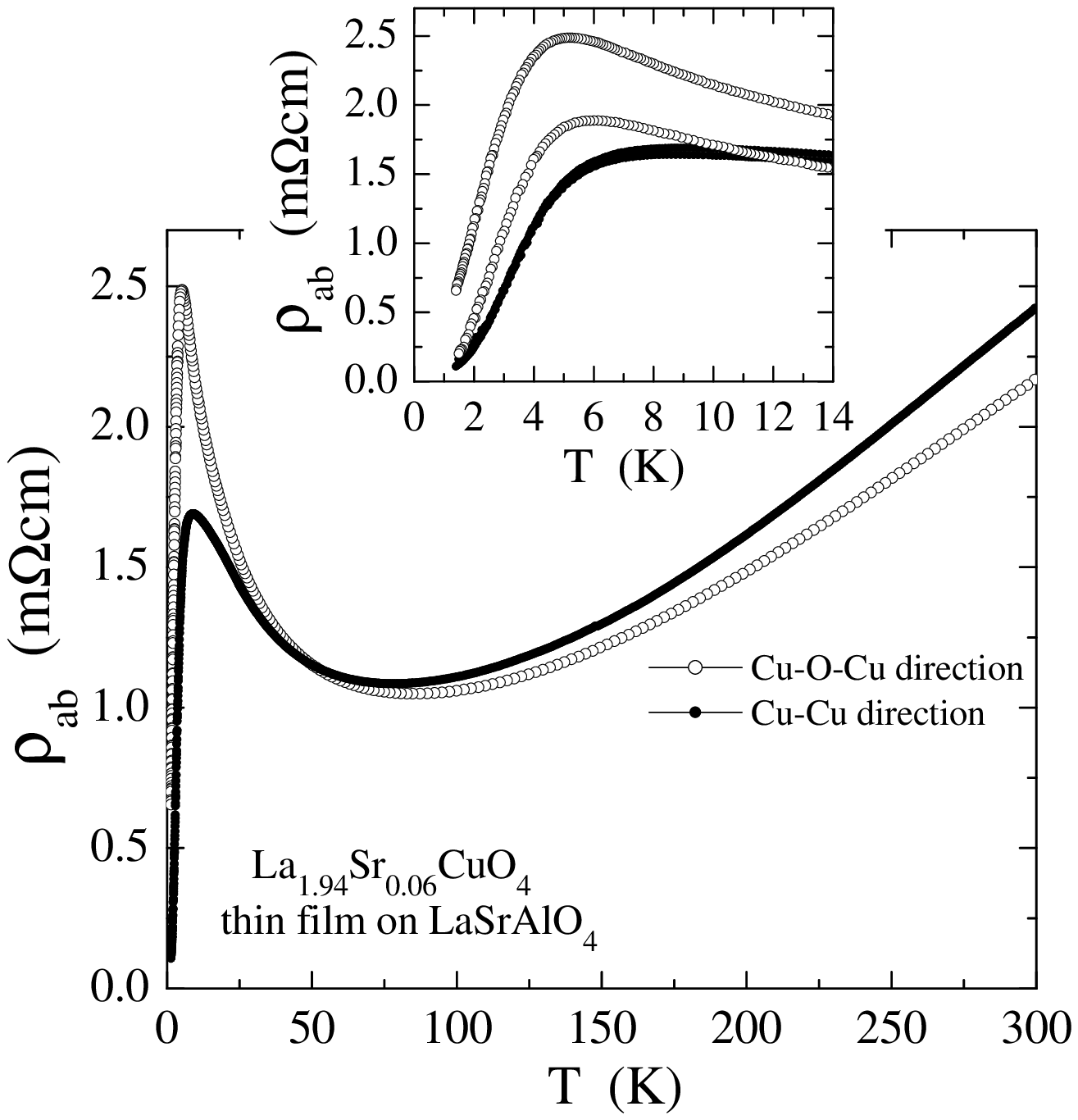}
\caption{Resistivity of 1000-\AA ~LSCO ($x=0.06$) films
deposited on LSAO; bridges are formed along the Cu-Cu or Cu-O-Cu
directions. Inset: resistivity of two pairs of bridges in the vicinity of
the superconducting transition.}
\label{R_LSAO}
\end{figure}

The low-temperature resistivity upturn in Figs. 2 - 4 reflects the process
of collective -- caused by the stripe pinning -- or individual
localization of holes. At low temperatures, neutron scattering also showed
the dynamic stripe correlations to slow down and to evolve into a static
order.\cite{diag_ver} Apparently, it should be this region where one may
expect high electric fields to overcome the pinning and to cause
non-linear conductivity features.

\subsection{Overheating effects}
\label{sec:heat}

Before proceeding to the $I-V$ measurements, let us first consider the
current-induced Joule heating, and estimate how high an electric field $E$
can be applied to a bridge without causing a significant increase of its
temperature. For the geometry of narrow bridges, where both the produced
heat and the heat removal scale with the bridge's length, the overheating
can be estimated rather easily, without complicated mathematics
\cite{hot_spot} required for bulk samples. Taking a typical thin (0.1
$\mu$m) bridge with a width of 25-50 $\mu$m, and assuming the substrate to
be thermally anchored at a distance of $\sim 1$ mm, one can calculate the
bridge overheating to be $\Delta T \approx 2P_l/\kappa_{\text{sub}}$,
where $P_l$ is a power being dissipated per unit length of the bridge and
$\kappa_{\text{sub}}$ is the thermal conductivity of the substrate. We can
therefore estimate the actual temperature for each bridge,
$T_{\text{br}}$, as a function of the applied electric field, using
experimental resistivity data $\rho_{ab}(T)$:
\begin{equation}
T_{\text{br}}(E)=T_{\text{base}}+\Delta T \approx
T_{\text{base}}+2E^2S[\rho_{ab}(T_{\text{br}})
\tilde{\kappa}_{\text{sub}}]^{-1},
\end{equation}
where $S$ is the bridge's cross-section, and $\tilde{\kappa}_{\text{sub}}$
is an effective heat conductivity in the range
$\kappa_{\text{sub}}(T_{\text{base}})$ to
$\kappa_{\text{sub}}(T_{\text{br}})$.

\begin{figure}[!tb]
\leftskip18pt
\includegraphics*[width=17pc]{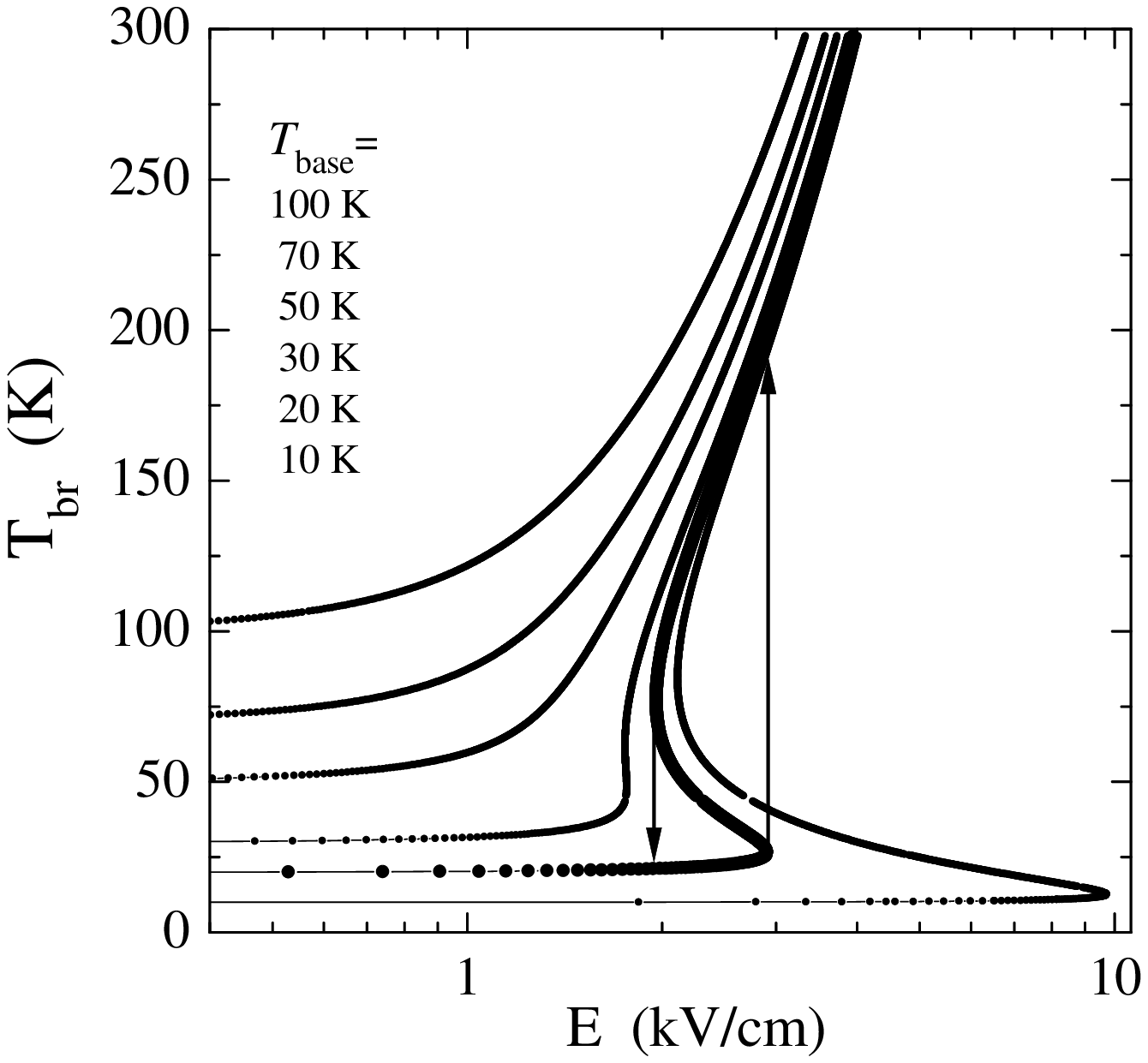}
\caption{Estimated temperature $T_{\text{br}}$ of a LSCO ($x=0.01$)
film bridge as a function of applied voltage for several base
temperatures; the heat conductivity of substrate $\kappa_{\text{sub}}$ is
taken as 150 mW/Kcm, cross-section of the bridge
-- 2.5 $\mu$m$^2$. Arrows indicate jumps that should occur upon increasing
and decreasing the electric field at the base temperature of 20 K.}
\label{overheat}
\end{figure}

Although in reality $\kappa_{\text{sub}}$ depends on the type of
substrates, and may vary strongly with temperature, a reasonable
qualitative picture of the overheating can be obtained by assuming
$\kappa_{\text{sub}}$ to have an average, temperature-independent value.
Figure 5 illustrates how an actual temperature of a typical LSCO $x=0.01$
bridge (0.1 $\mu$m thick, 25 $\mu$m wide) should change with applied
electric field; the calculations are done using the experimental
$\rho_{ab}(T)$ data and taking $\kappa_{\text{sub}}\sim 150$ mW/Kcm.
Apparently, as the applied electric field reaches several hundreds V/cm,
the actual temperature of the bridge should deviate considerably from the
base temperature; this deviation is stronger at higher $T_{\text{base}}$,
where the bridge conductivity and thus the produced power are larger. For
low base temperatures, the smooth heating becomes unstable because of a
positive feedback: As the bridge is heated, its resistivity drops and the
produced power grows much quicker than the heat removal does, causing a
thermal instability and a very abrupt increase in temperature by several
hundreds degrees (Fig. 5). In fact, for realistic $\kappa_{\text{sub}}(T)$
that decreases at high temperatures, the high-$T$ branches of the curves
in Fig. 5 become almost vertical, so that an applied voltage of several
kV/cm would literally burn the sample.

The non-linearity in $I-V$ characteristics, that follows from the
calculated Joule overheating, is shown in Fig. 6. It turns out that the
differential resistance, $dV/dI$, may stay virtually unchanged up to the
electric field $\sim 0.1-0.3$ kV/cm, but it should show a spectacular drop
upon further increasing the voltage, as the bridge gets heated by the
current. The arrows in Fig. 6 indicate an inevitable switching between the
high and low-resistance states accompanied by a hysteresis -- the
phenomena that are unrelated to any electronic peculiarities, but are
governed exclusively by the conventional heating. Upon measuring the $I-V$
characteristics and interpreting the data we therefore should keep in mind
the threshold field of $0.1-1$ kV/cm, where the Joule heating becomes
crucial.

\begin{figure}[!tb]
\includegraphics*[width=20.5pc]{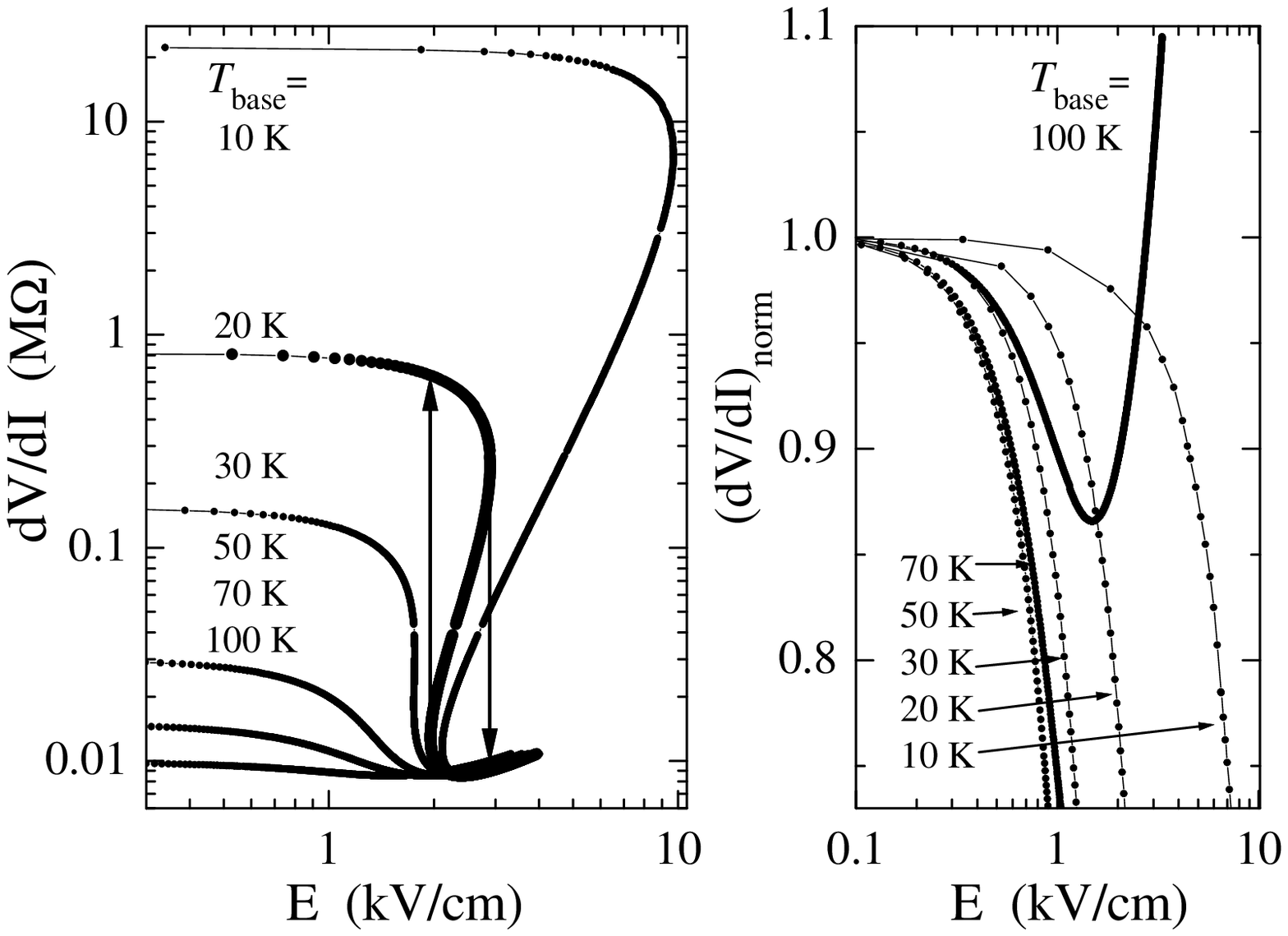}
\caption{(Left) Electric-field dependence of the differential resistance
$dV/dI$ that should be caused solely by the bridge overheating, as
estimated in Fig. \ref{overheat}. Arrows indicate jumps that should occur
upon increasing and decreasing the electric field at the base temperature
of 20 K. (Right) Differential resistance normalized to its low-field
value.}
\label{IVCh}
\end{figure}

One may wonder whether a pulse technique can be helpful in avoiding the
overheating problem; thus, it is instructive to estimate the
characteristic time for the sample heating. For example, at $T=20$ K the
heat capacity per unit length of a bridge (with a cross section of 2.5
$\mu$m$^2$) can be estimated as $C_l(20K)\sim 2.5$ $\mu$m$^2\times0.1$
J/Kcm$^3$=$2.5\times10^{-9}$ J/Kcm. When an electric field of 1 kV/cm is
applied to the bridge at 20 K [Fig. 6(a)], the produced power is
$P_l\approx 12$ mW/cm (or merely $\sim 0.1$ mW for our 100-$\mu$m-long
bridge). In an equilibrium state, when the heat is removed through the
substrate with $\kappa_{\text{sub}} \sim 150$ mW/Kcm, the power $P_l=12$
mW/cm would cause just a minor overheating by $\Delta T \approx
2P_l/\kappa \approx 0.16$ K. However, in the absence of heat removal, this
seemingly small power would heat the bridge at a rate of $dT/dt = P_l/C_l
\sim 5\times 10^6$ K/s; apparently, the bridge's temperature should approach
its equilibrium value within an extremely short time of $\sim 0.1$ $\mu$s.
In the case of thin films, the heat capacity is therefore a poor
competitor to the heat conductivity in controlling the overheating rate,
and thus the pulse technique can hardly be helpful.

It should be noticed that the above estimates are done for LSCO $x=0.01$,
while for samples with higher doping the ``safe'' electric field decreases
as a square root of the resistivity; it also decreases with increasing the
sample's cross section, so that bulk samples can be significantly
overheated by orders of magnitude smaller electric fields.

\subsection{Current-voltage characteristics}

\begin{figure}[!t]
\includegraphics*[width=17.5pc]{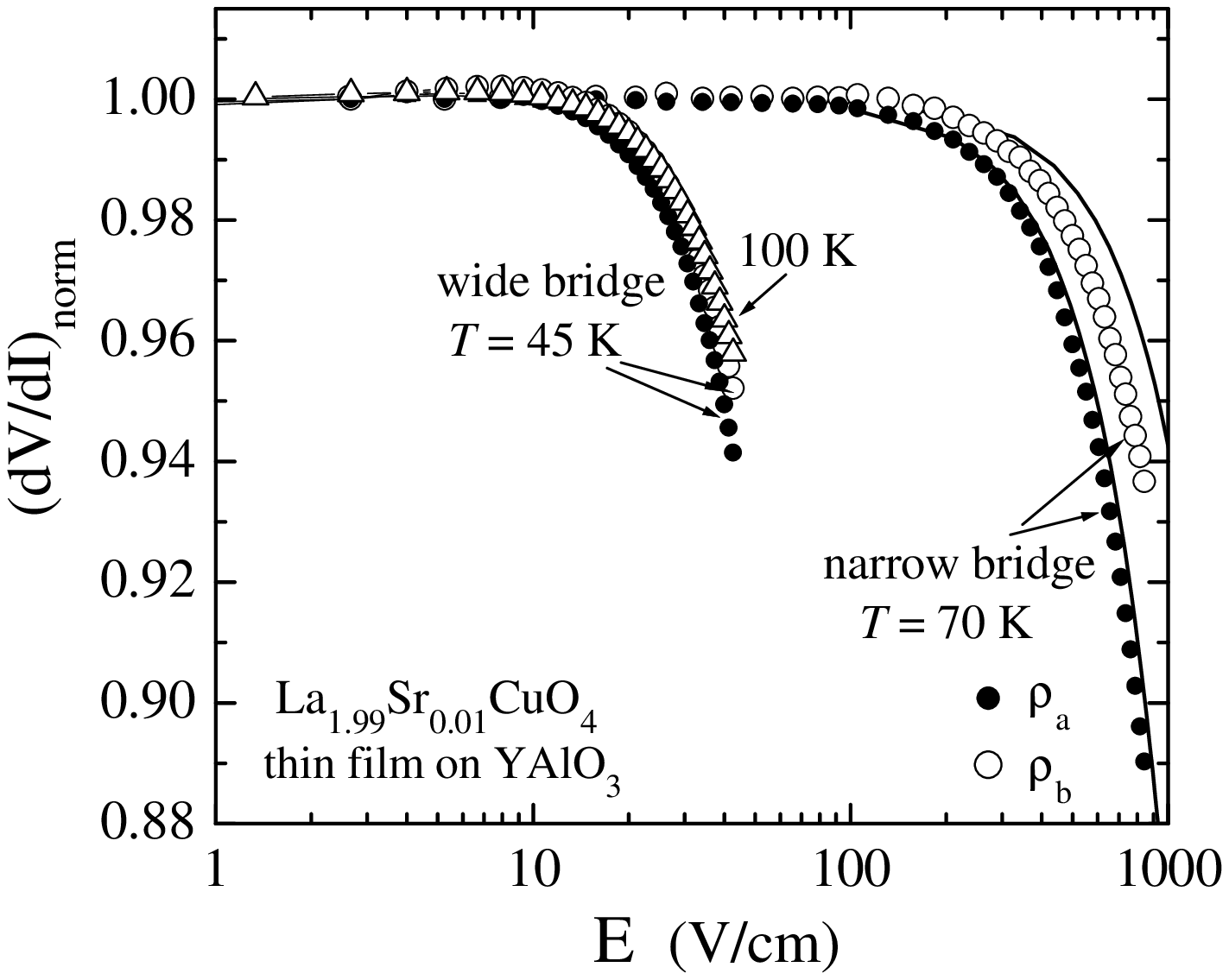}
\caption{Normalized differential resistance of LSCO ($x=0.01$)
thin-film bridges as a function of dc bias field. The presented data were
taken at 45 K (circles) and 100 K (triangles) on wide, $460-500 \mu$m,
bridges (1200-\AA ~films); and at 70 K on narrow, $18-20 \mu$m bridges
(2400-\AA ~films). Solid and open symbols show the resistance measured
along the $a$ and $b$ axes, respectively. Solid lines indicate an
estimated effect of overheating for the narrow bridges.}
\label{IV_YAP}
\end{figure}

Upon looking for non-linear conductivity features related to the
collective charge motion, we have measured the differential resistance
$dV/dI$ of narrow ($\sim 20$ $\mu$m) LSCO $x=0.01$ bridges by sweeping the
bias field up to $\sim 1$ kV/cm. Measurements were performed at fixed
temperatures in the range from 150 K (where $\rho_{ab}(T)$ has a minimum,
see Fig. 2) down to 40 K (where $\rho_{ab}$ exceeds the minimum value by
several times); typical $dV/dI$ data taken at $T=70$ K are shown in Fig.
7. The $I-V$ characteristics turn out to be perfectly linear, and thus the
differential resistance stays unchanged, up to rather high fields of $\sim
100$ V/cm. Upon further increasing the voltage, the differential
resistance goes down, dropping by $\sim 10$\% as the field approaches 1
kV/cm. However, this resistivity decrease is smooth, without any step-like
feature that one would expect for the collective stripe sliding; moreover,
it well fits the overheating effect estimated for each bridge using its
resistivity $\rho_{ab}(T)$ and the heat conductivity of the substrate
(solid lines in Fig. 7). In order to confirm that the $I-V$ non-linearity
emerging at high voltages is caused solely by the Joule heating, we have
measured several bridges with different geometries. Since the produced
heat scales with the sample's volume, while the heat removal rate changes
rather slowly, the onset of non-linearity in larger bridges should take
place at lower electric fields. Figure 7 demonstrates that this is indeed
the case: Wide bridges show non-linearity starting already at 10-20 V/cm.
Apparently, the obtained data leave little room for any {\it intrinsic}
non-linear conductivity in LSCO $x=0.01$ films, at least at moderate
electric fields that do not cause significant overheating.

The $I-V$ characteristics measured on LSCO ($x=0.06$) bridges also show a
perfectly linear behavior up to electric fields of $20-30$ V/cm (Figs. 8,
9), that is, as long as the field stays within the ``safe'' range where
the estimated Joule heating is negligible. At higher fields, the $dV/dI$
data deviate from a constant value, however this deviation clearly traces
the temperature dependence of $\rho_{ab}$, giving an additional evidence
for the overheating mechanism. For example, $dV/dI(V)$ dependences
measured for LSCO ($x=0.06$) bridges at $T=68$ K -- somewhat below the
resistivity minimum -- exhibit a non-monotonic behavior, also passing
through a minimum (Fig. 8). In fact, what we see in the $dV/dI(V)$ curves
is simply an increase of the actual bridge's temperature $T_{\text{ch}}$,
so that $dV/dI$ is equal to the resistivity $\rho_{ab}(T_{\text{ch}})$. A
good quantitative agreement of the data with the fit in Fig. 8 clearly
indicates that there is no other source for the non-linear conduction,
besides overheating.

\begin{figure}[!t]
\includegraphics*[width=17.8pc]{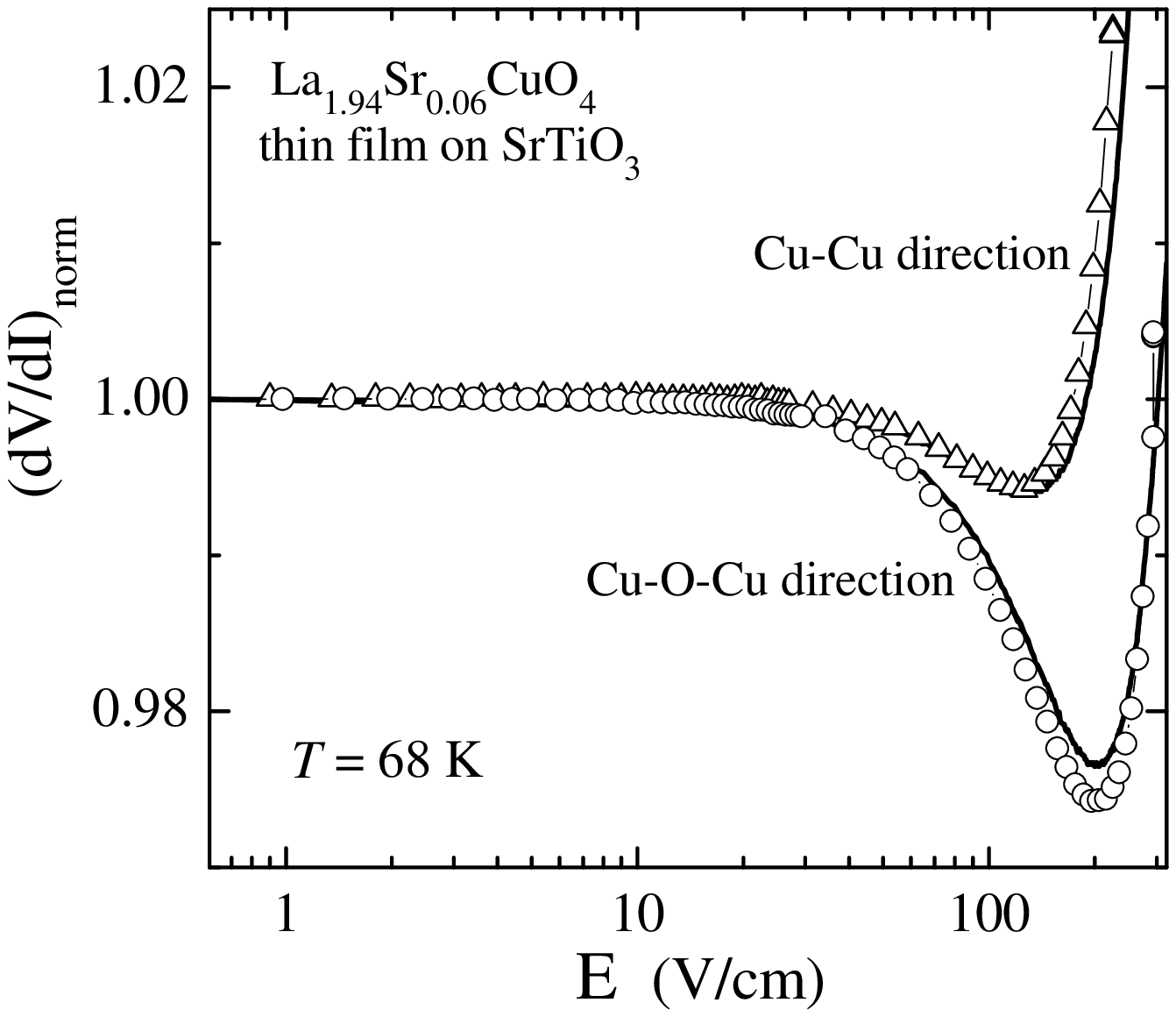}
\caption{Normalized differential resistance of LSCO ($x=0.06$)
films measured at $T=68$ K as a function of dc bias field. Solid lines
show an estimated effect of the bridge overheating.}
\label{IV_STO}
\end{figure}

\begin{figure}[!tb]
\includegraphics*[width=17.6 pc]{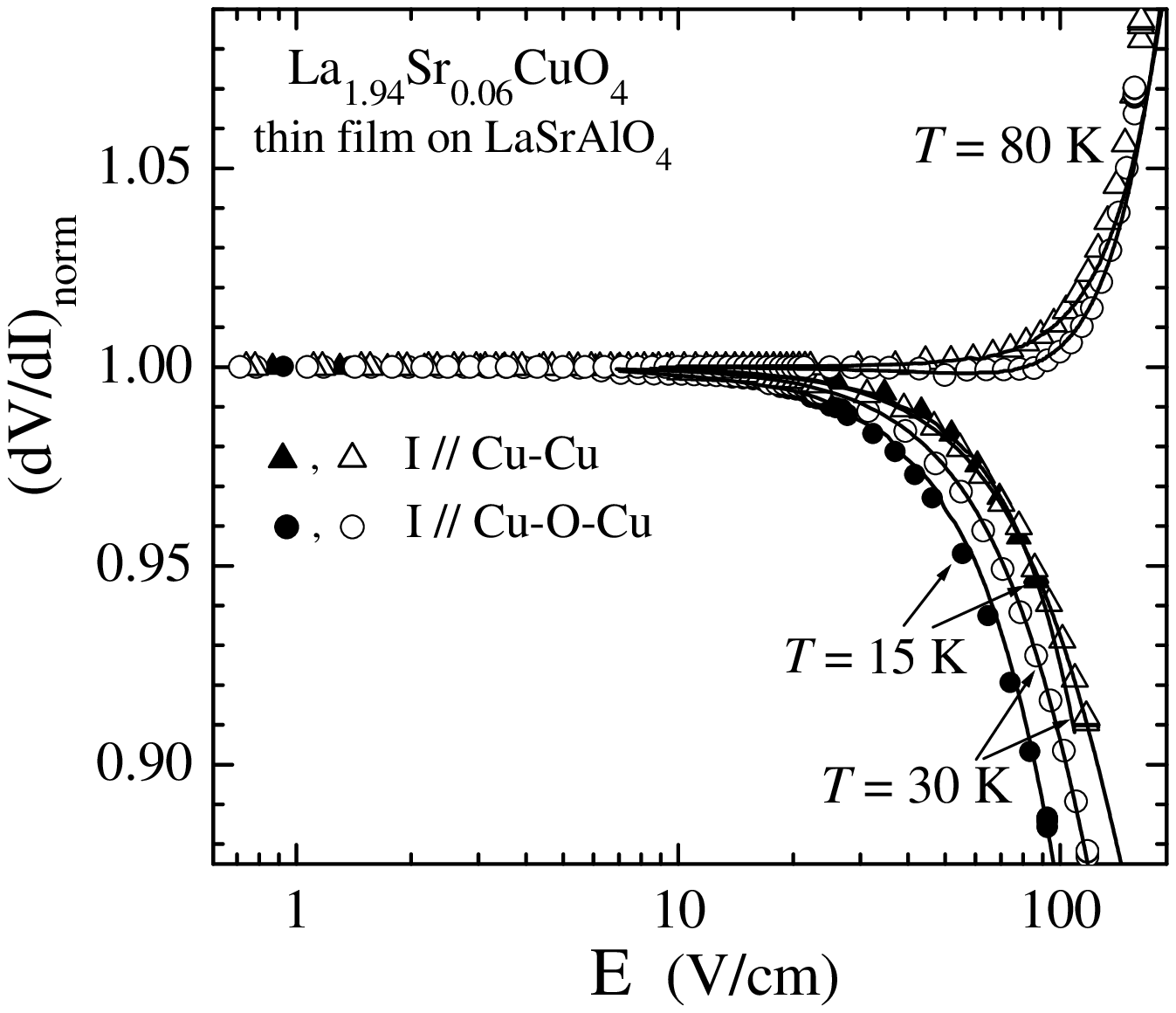}
\caption{Normalized differential resistance of LSCO ($x=0.06$)
films at several temperatures as a function of dc bias field. Solid lines
show the estimated effect of overheating.}
\label{IV_LSAO}
\end{figure}

Figure 9 presents the $dV/dI$ data obtained for LSCO ($x=0.06$) bridges
deposited on LaSrAlO$_4$. Depending on whether the measurements are done
at temperatures where $d\rho_{ab}/dT$ is positive or negative (see Fig.
4), the differential resistance increases or decreases with increasing
electric field, exactly as expected for the non-linearity originating
exclusively from the Joule overheating. No other features could be
detected in the $I-V$ curves at any temperature down to the onset of
superconductivity at $T<10$ K.

To summarize the experimental observations, we can state that no signs of
{\it intrinsic} non-linear conductivity are found in LSCO ($x=0.01$ and
0.06) thin films when electric fields up to several hundreds V/cm are
applied along any crystallographic direction.

\subsection{Do the stripes actually exist in cuprates?}

Since the performed experiments could not reveal any non-linear feature
related to the collective charge motion, a natural question to be asked is
whether this negative result can somehow be reconciled with the existence
of charged stripes. In fact, the only obvious possibilities are that the
charged stripes in LSCO, if actually exist, are either pinned so strongly
that available electric fields appear to be too weak to induce their
sliding, or they are instead not pinned at all and exhibit a linear
fluid-like behavior even at the lowest fields. The latter possibility,
however, sounds quite unlikely given the insulating tendency of the
resistivity at low temperatures (Fig. 2). We should therefore consider the
conditions that may prevent the charge order from being dragged by
electric fields; then the limitations imposed by the present result on the
picture of stripes in cuprates will become clear.

In general, the electrical conductivity of solids becomes non-linear when
electrons accelerated by an applied electric field $E$ acquire an energy
$eEl$ (where $l$ is the hopping distance or mean free path) comparable to
other relevant energy scales such as the Fermi energy $\varepsilon_F$, the
band gap $\Delta$ , or $k_BT$; usually this occurs at very high fields,
$\sim 10^4-10^7$ V/cm. What is specific to charge-ordered systems is that
the electric-field effect is integrated over a macroscopic number of
electrons being able to move cooperatively. Consequently, the
characteristic fields are reduced dramatically, roughly speaking by as
many times as the number of electrons involved in the cooperative motion.
The observation of a threshold conductivity at small fields thus implies
that the following conditions are met: (i) the charge order is stiff
enough to keep its phase over a fairly large coherent domain, whose volume
$V=L_x \times L_y \times L_z$ contains $N_e=Vn_e\gg 1$ electrons
participating in the CDW; (ii) pinning of such domain by the lattice or
impurities is substantially stronger than thermal fluctuations, $k_BT$,
otherwise the system exhibits a fluid behavior without any threshold for
conduction; (iii) a force exerted by a fairly small electric field on a
phase-correlated domain, $eN_eE$, can overcome the pinning. In fact, the
latter two conditions are also related to the CDW stiffness: Thermal
fluctuations become irrelevant for macroscopic domain sizes, and a stiffer
CDW is pinned less readily by {\it uncorrelated} defects. \cite{CDW_rev,
cuprates, CDW_theo}

In inorganic chain compounds such as transition-metal chalcogenides
NbSe$_3$, TaS$_3$, or blue bronze K$_{0.3}$MoO$_3$, the key to spectacular
non-linear conductivity phenomena is an extremely large coherence length
of the charge order, reaching $\sim 1-100$ $\mu$m. \cite{CDW_rev,
CDW_size, CDW_dom} Correspondingly, the threshold field for depinning the
CDW is reduced by many orders of magnitude from characteristic
single-electron values down to 1-100 mV/cm, \cite{CDW_rev, CDW_thr,
CDW_size, commen} and the features related to the CDW sliding stay sharp
up to the CDW-formation temperature, since thermal fluctuations have no
impact on macroscopic correlated domains.

The fact that electric fields of several hundreds V/cm are unable to
induce non-linear conduction in lightly doped LSCO indicates that the
phase-correlated domains here should be much smaller than in chain
compounds. Let us roughly estimate how small they should be. According to
neutron scattering, the stripes become static at temperatures below 10-30
K, implying the pinning energy per domain to be of the order of several
meV. By comparing the work that an electric field would do upon dragging
stripes by one lattice constant $a$, $eN_eEa$, with the pinning energy,
one can estimate that electric fields $E\sim 1$ kV/cm would be incapable
of depinning phase-correlated domains if they contain $N_e \leq 100$
electrons. More sophisticated calculations by Morais Smith {\it et al.}
\cite{cuprates} predict $N_e \sim 100$ and a stripe-depinning field in
LSCO $x=0.01$, $E_c \sim 10^4$ V/cm. Whatever the case, the phase
coherence in LSCO can hardly exceed $\sim 100$ lattice constants along the
direction of stripes and more than just a few periods in transverse
directions. It is worth noting however that the above estimates do not
imply the stripes to be fragmented, they only indicate the length scale
over which the stripe structure can behave as a {\it stiff} object.

\begin{figure}[!t]
\includegraphics*[width=20 pc]{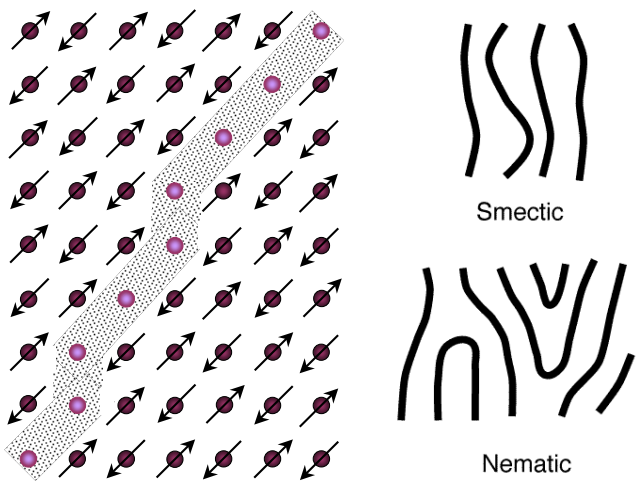}
\caption{(Left) A charge stripe separating antiferromagnetic domains in a
CuO$_2$ plane; arrows indicate spins localized on Cu ions. The stripes at
low doping are believed to be essentially diagonal, yet they can easily
contain kinks or vertical fragments. The vertical stripes at higher
doping, in turn, may include diagonal parts. (Right) Possible topologies
of stripes suggested in Ref.~[\onlinecite{Kivelson}].}
\label{stripes}
\end{figure}

Apparently, the charge stripes in cuprates with so short coherence length
should look like a ``spaghetti'' of flexible weakly-interacting strings
(Fig. 10), rather than a conventional rigid CDW. The term ``electronic
liquid crystal'' has been coined to describe such unusual state of matter.
\cite{Kivelson} What, however, makes the stripes so different from CDW,
and what allows them to be flexible? In 1D chain compounds, the stiffness
of CDW comes from its insulating nature: The electronic energy is reduced
owing to a gap opening at the Fermi energy, and the CDW period is strictly
determined by the Fermi wave number $k_F$. Consequently, any forced
modification of the CDW period would shift the gap away from the Fermi
surface, and thus inevitably destroy the CDW. In 2D systems, the metallic
state usually survives the CDW formation,\cite{CDW_2D} and the resulting
CDW is much less stiff: it can modify the periodicity and orientation to
fit different parts of the Fermi surface. The charged stripes in cuprates
have a good reason to be even more flexible; they are formed because holes
are expelled from antiferromagnetic domains, while a possible ordering of
stripes into a periodic structure is merely a secondary effect. In fact,
the stripe's flexibility originates from the absence of a gap at the Fermi
level, which allows the stripes to change their filling (number of holes
per unit length) and thus the spacing between adjacent stripes. Moreover,
the energy of stripes only weakly depends on their orientation within
CuO$_2$ planes, since both vertical and diagonal ones are experimentally
observed \cite{diag_ver} in lightly doped LSCO. This makes it easy for
stripes to bend and form kinks, as is sketched in Fig. 10. Needless to say
that such ``electronic liquid crystals'', where each small fragment slides
virtually independently and can adjust itself to the ionic lattice or
impurities, should never exhibit any threshold conduction features. In
this sense, our observation of linear $I-V$ characteristics gives evidence
that if the charge-density modulations actually exist in underdoped
cuprates they should be of the electronic liquid-crystal type.

\subsection{Implications for other transition-metal oxides}

One might wonder whether the absence of threshold conduction in LSCO
indicates a fundamental difference of the stripes in cuprates from the
charge ordered states in other transition-metal oxides, such as nickelates
or manganites. \cite{Nature,Science,LaSrNiO,film,film_perp} This may
indeed be true, since the charge-ordered nickelates or manganites usually
possess much higher resistivity than the layered cuprates do. However, a
close analysis of relevant publications shows that there is, in fact, no
unambiguous evidence for the collective charge motion in other layered (or
3D) oxides either. The spectacular resistivity switching found in single
crystals of manganites, nickelates, or chain cuprates SrCuO$_2$ and
Sr$_2$CuO$_3$ always takes place at remarkably similar
conditions,\cite{Nature, Science, LaSrNiO,SrCuO} implying that
peculiarities of the electronic and crystal structures of these compounds
may not be the key for this phenomenon. Moreover, the observed
characteristic threshold field $E_{\text{th}}$ of the order of several
kV/cm, and the shape of the $I-V$ characteristics, both are very similar
to what should be expected for the heating effects (Fig. 6). In all these
experiments, the power dissipation in the low-resistance state was $\sim
100$ mW; given a rather small heat conductivity of these samples (for
example, in manganites \cite{kappa_Mn} $\kappa \sim 30$ mW/Kcm), this was
large enough to overheat the mm-size crystals by $>10$ K, let alone a much
stronger local heating possible for inhomogeneous current flows.
\cite{Bonch, hot_spot}

Of course, it would be incorrect to attribute the non-linear conductivity
in manganites and nickelates entirely to the Joule heating: electronic
inhomogeneities inherent in these compounds may set an arrangement of
conducting filaments, the flowing current may alter the charge-order
domain structure, etc. The problem is, however, that the electric fields
required to induce the resistivity switching in layered oxides are clearly
out of the ``safe'' range, and thus a special care should be paid to
distinguish an initial coherent charge-order sliding, \cite{LaSrNiO} if it
actually takes place, from the following heating effects that quickly mess
everything up. Such problems were often encountered upon $I-V$
measurements of 2D electron systems \cite{2DEG_heat} and semiconductors
\cite{hot_spot,not_ther} as well.

When a high electric field is applied to an insulating sample, a
homogeneous current distribution becomes unstable, \cite{Bonch} and a kind
of ``spark'' may develop along the best conducting path, tending to spread
and burn the sample. If, however, the total current flowing through the
sample is limited, the spark channel optimizes its size to keep the
temperature high enough for providing the required conductivity. A
self-optimized channel may collect virtually all the current flowing
through the sample,\cite {hot_spot,not_ther} rising its density up to
$10^3-10^5$ A/cm$^2$. \cite{density} In a sense, such conducting filament
inside a crystal is quite reminiscent of our conducting bridges on
insulating substrates. As we discussed in Sec.\ \ref{sec:heat}, a power of
$\sim 1$ mW (typical for high-resistivity state \cite{SrCuO, LaSrNiO}) can
considerably overheat a bridge, or equivalently, a conducting filament
within several microseconds; apparently, the experimentally observed
\cite{SrCuO, LaSrNiO} switching delays of 1-1000 ms provide more than
enough time for the heating process to develop.

To understand whether or not the temperature of conducting filaments
actually exceeds significantly the average temperature of crystals, one
needs to know the exact geometry of filaments. \cite{hot_spot,not_ther} An
optical study of Pr$_{0.7}$Ca$_{0.3}$MnO$_3$ crystals has shown that a
0.15-mm-long conducting filament expands up to 0.2 mm in diameter as the
dissipating power reaches $\approx 90$ mW. \cite{Science} For that
particular filament geometry and the heat conductivity $\kappa
\sim 30$ mW/Kcm (Ref. [\onlinecite{kappa_Mn}]), one can estimate that the
overheating {\it must} be rather large, $\sim 50-100$ K, which alone can
induce a resistivity switch. The role of heating becomes more clear when
samples with different sizes are compared: In Pr$_{0.7}$Ca$_{0.3}$MnO$_3$
thin films, the resistivity switching has been found \cite{film_perp} to
occur at significantly higher fields, $E_{\text{th}} \approx 2 \times
10^5$ V/cm; interestingly, the power dissipation in the low-resistance
state of these films still appears to be virtually the same as in single
crystals, $\sim 100$ mW. This power released on the surface of 2500
$\mu$m$^2$ -- exactly as the surface of our bridges (25 $\mu$m $\times$
100 $\mu$m) -- had to overheat the film by $\geq 100$ K, which well
accounts for the observed resistivity drop.

It turns out, therefore, that the non-linear conductivity in layered
transition-metal oxides is observed only at very high electric fields
where heating effects should become crucial. Consequently, thus far one
has insufficient information to conclude whether or not a coherent sliding
of stripes (or another charge order) can ever be induced in these
compounds. It may well be that the charge ordering in layered oxides is
always of the same kind -- flexible and readily adjustable to the ionic
lattice and impurities.

\section{Conclusions}

The ubiquity and properties of the charged stripes in high-$T_c$ cuprates
still remain an issue. We have tried to induce a coherent sliding of the
charged stripes in La$_{2-x}$Sr$_x$CuO$_4$ ($x=0.01$ and 0.06) thin films
by applying high electric fields up to 100-1000 V/cm, yet observed no
non-linear conductivity features, at least as long as the films are not
overheated significantly by the flowing current. This result can be
reconciled with the existence of charged stripes only if they are very
flexible, since the less stiff order is known to be pinned more readily.
Simple estimates show that the volume over which the stripes move
coherently can hardly include more than $\sim 100$ holes, implying the
stripe fragments are capable of moving virtually independently.
Consequently, the self-organized electronic structures in cuprates, and
presumably in other layered oxides, should be considered as a kind of
``electronic liquid crystal'' rather than as a superposition of rigid
charge and spin density waves.

\begin{acknowledgments}
We thank K. Segawa for invaluable technical assistance.
\end{acknowledgments}


\begin{thebibliography}{99}
\bibitem{density} Note, however, that the current density inside
conducting filaments can never be as large as $10^{11}$ A/cm$^2$ as was
suggested in Ref.~[\onlinecite{EtoH}] to explain the non-linear $I-V$
behavior in manganites, and thus cannot generate any sizeable magnetic
field.
\end{thebibliography}
\end{document}